\documentclass{aa}

\usepackage{amsmath}
\usepackage{amssymb}
\usepackage{latexsym}
\usepackage{graphicx}
\usepackage[varg]{txfonts}
\usepackage{natbib}
\usepackage{color}
\bibpunct{(}{)}{;}{a}{}{,} 

\usepackage{pdflscape}
\usepackage{afterpage}

\usepackage{amstext}
\usepackage{lipsum}
\usepackage{subfig}

\newcommand{\kms}{km\,s$^{-1}$}

\begin{document}

\title{Dynamics of internetwork chromospheric fibrils: Basic properties and MHD kink waves}

\author{K. Mooroogen$^1$, R. J. Morton$^1$, V. Henriques$^{2,3}$} 

\institute{$^1$ Department of Mathematics, Physics and Electrical Engineering, Northumbria University, Ellison Building,
Newcastle upon Tyne, NE1 8ST, UK, \email{richard.morton@northumbria.ac.uk}\\
$^2$ School of Mathematics and Physics, Queen’s University Belfast, Belfast, Northern Ireland, UK\\
$^3$ Institute of Theoretical Astrophysics, University of Oslo, P.O. Box 1029, Blindern, N-0135 Oslo Norway}

\abstract{}
{Current observational instruments are now providing data with the necessary temporal and spatial cadences required to examine highly 
dynamic, fine-scale magnetic structures in the solar atmosphere. Using the spectroscopic imaging capabilities of the 
Swedish Solar Telescope, we aim to provide the first investigation on the nature and dynamics of 
elongated absorption features (fibrils) observed in H$\alpha$ in the internetwork. } 
{We observe and identify a number of internetwork fibrils, which form away from the kilogauss, network magnetic flux, 
and we provide a synoptic view on their behaviour. 
The internetwork fibrils are found to support wave-like 
behaviour, which we interpret as Magnetohydrodynamic (MHD) kink waves. The properties of these 
waves, that is, amplitude, period, and propagation speed, are measured from time-distance diagrams and we
attempt to exploit them via magneto-seismology in order to probe the variation of plasma 
properties along the wave-guides.}
{We found that the Internetwork (IN) fibrils appear, disappear, and re-appear on timescales of tens of minutes, suggesting
that they are subject to repeated heating. No clear photospheric footpoints for the fibrils are found in
photospheric magnetograms or H$\alpha$ wing images. However, we suggest that they are magnetised features as the majority
of them show evidence of supporting propagating MHD kink waves, with a modal period of $120$~s. 
Additionally, one IN fibril is seen to support 
a flow directed along its elongated axis, suggesting a guiding field. The wave motions are found to propagate at 
speeds significantly greater than estimates for typical chromospheric sound speeds. {Through their interpretation 
as kink waves}, the measured speeds provide an  
estimate for local average Alfv\'en speeds. Furthermore, the amplitudes of the waves are also found to vary as a function of distance
along the fibrils, which can be interpreted as evidence of stratification of the plasma in the neighbourhood
of the IN fibril.}
{} 

\date{Received /Accepted}

\titlerunning{IN chromospheric fibrils}
\authorrunning{Mooroogen et al.}

\maketitle

\section{Introduction} \label{sec:intro}
Our ability to probe the solar chromosphere has seen significant advances in the recent
past. The increased spatial, temporal, and spectral resolution of both ground and space based
observatories has provided unique insights into the highly dynamic behaviour of fine-scale
features. The observation of energy transfer by Magnetohydrodynamic (MHD) waves is one area that 
has benefited from these improvements, with the identification of ubiquitous transverse waves --
at least in chromospheric features associated with enhanced photospheric magnetic field 
concentrations that emanate from the magnetic network and plage, that is, spicules 
(e.g. \citealp{DEPetal2007}) and fibrils (e.g. \citealp{PIEetal2011}, \citealp{MORetal2012c}). 

Estimates for the energy flux imply 
that the waves carry enough energy for plasma heating. However, current techniques for estimating the energy 
flux are generally crude, meaning that the actual energy content of the waves and their role in
atmospheric plasma heating is subject to uncertainty (see e.g. \citealp{GOOetal2013}, \citealp{VANetal2014}). 
There is further uncertainty surrounding the mechanism(s) by which 
these waves can actually deposit their energy in the local plasma. By their very nature they are incompressible
and difficult to dissipate. Currently, the most favoured mechanism for the damping of the kink mode is
resonant absorption (\citealp{TERetal2010c}, \citealp{VERTHetal2010}), although this process is just a 
transfer of energy from the kink modes to the quasi-torsional
$m=1$ Alfv\'en modes, and requires a further mechanism to dissipate their energy, for example, phase mixing of the $m=1$ Alfv\'en modes (\citealp{SOLetal2015}). An alternative mechanism 
may involve the generation of instabilities at the boundaries of 
the flux tube by the kink modes (\citealp{TERetal2008b}, \citealp{ANTetal2015}). In order to develop our understanding,
detailed observations and analysis of MHD waves in the solar atmosphere are required, and one of the key objectives will
be to determine the role the chromosphere plays in regulating the energy flow in the atmosphere.

In general, recent studies of chromospheric phenomena have focused on events that occur around network concentrations
of magnetic flux and plage regions, with a wealth of individual features identified (e.g. Rapid blue-shifted events -RBEs, Type I and II spicules; \citealp{ROUetal2009}, \citealp{PERetal2012}).
This focus is primarily due to the magnetic connection of these features to the corona and the hypothesis that they
contribute to the mass and heat flux required to sustain the corona (e.g. \citealp{DEPetal2011}, \citealp{PERetal2014},
\citealp{HENetal2016}). In contrast, there has been much less attention given to the study of fibrils, 
which reach out from the network across 
the internetwork (IN), forming dense canopies that occupy a large volume of the visible 
chromosphere\footnote{Here we use the term chromosphere to refer to {the traditional definition 
of the region of the Sun's atmosphere} which is observed during eclipses (\citealp{RUT2010}).} (\citealp{RUT2006}). In addition to the 
network fibrils there exist shorter, absorbing H$\alpha$ features that appear in the IN, which have gone somewhat 
unmentioned in literature (a discussion of these features is the subject of this paper). 

Observations reveal that the fibrils display time-dependent behaviour, appearing,
disappearing, and reappearing over tens of minutes, likely indicating a departure from a hydrostatic plasma. Such
a thermodynamic cycle must then require an additional and time-dependent energy input into the system. It has been
suggested that the fibrils are the signatures of heating events or at least the markers that heating events may have 
occurred at an earlier, recent instance in time (\citealp{RUTVAN2016}). However, there is no consensus 
about the processes that lead to the formation of fibrils. Therefore, it seems prudent to engage in detailed
studies of the fibrils using the improved capabilities of modern instruments to gain insight into their nature,
and, as such, shedding light on the heating of the chromosphere. 

\medskip

In anticipation of later discussion here, we summarise some salient features of IN magnetism here. 
On average the IN magnetic flux makes up 14~\% of the total Quiet Sun flux and appears to 
migrate towards the network boundaries (potentially due to supergranule flows), providing a substantial contribution to 
sustaining the network flux through mergers (e.g. \citealp{WANZIR1988}; \citealp{WANetal1995}; \citealp{GOSetal2014}). Typical 
estimates of
photospheric IN magnetic field strengths are on the order of $\sim200$~G but there is evidence for the existence of kGauss features 
(\citealp{DOMetal2003}; \citealp{OROBEL2012}). The IN elements have lifetimes on the order of minutes but some survive longer in order to contribute to the network \citep{ZOUetal2010}. Extrapolations that include the IN field {demonstrate that short, closed magnetic fields }
should be ubiquitous (\citealp{SCHTIT2003}; \citealp{WIEetal2010}), with only a small fraction reaching
the corona.
\cite{WIEetal2010} suggest photospheric field strengths above 300~G contribute to around 90~\% of the chromospheric magnetic
energy. Analogous low-lying magnetic fields have potentially been identified in recent high spectral resolution Ca II H data 
of active regions (\citealp{JAFetal2016b}).

The magnetic fields in the photosphere are also observed to be highly dynamic. Study of the motions of magnetic bright points 
in G-band images (\citealp{BERTIT1996}; \citealp{NISetal2003}, \citealp{CHITetal2012}) and flux concentrations in magnetograms 
(\citealp{GIAetal2014a,GIAetal2014}) has revealed that 
these small-scale magnetic elements migrate through the photosphere, whether due to some local flow or larger supergranule
flow pattern. Superimposed on the longer timescale migration are random motions of shorter timescales due to granular buffeting
(e.g. \citealp{DEWetal2005}). 
It is assumed that such motions can excite MHD waves, in particular transverse waves (e.g. \citealp{CHOetal1993}). 
\cite{STAetal2015} provided evidence
that these motions excite kink waves in chromospheric network bright points in Ca II H (3969~{\AA}). \cite{MORetal2013,
MORetal2013b} also demonstrate a correspondence between photospheric bright point motions and the transverse waves of 
long network fibrils in H$\alpha$ (cf \citealp{HILetal2013} for prominences). There is no evident reason to believe that 
photospheric velocity fields cannot also excite waves along IN magnetic fields.

In this study, we begin to examine the nature and dynamics of chromospheric features located in the IN, focusing in particular on the
apparent periodic transverse displacement of the IN fibrils (see Section \ref{sec:inf_ds} for definition and description). The periodic 
displacements of the IN fibrils axis are interpreted in terms of the kink wave. The data used here also has a
very high temporal cadence ($\sim$1~s), which provides an ideal opportunity to measure the range of propagation speeds of 
kink waves in the chromosphere, something that has proved difficult to do in previous studies due to 
the large Alfv\'en speed and short-scales of the features (\citealp{JESetal2015}). Furthermore, we exploit
measured wave properties to investigate the nature of the IN fibrils.

\begin{figure*}
        \centering
        \includegraphics[scale=0.9, clip=true, viewport=0.5cm 10.4cm 22.cm 17.5cm]{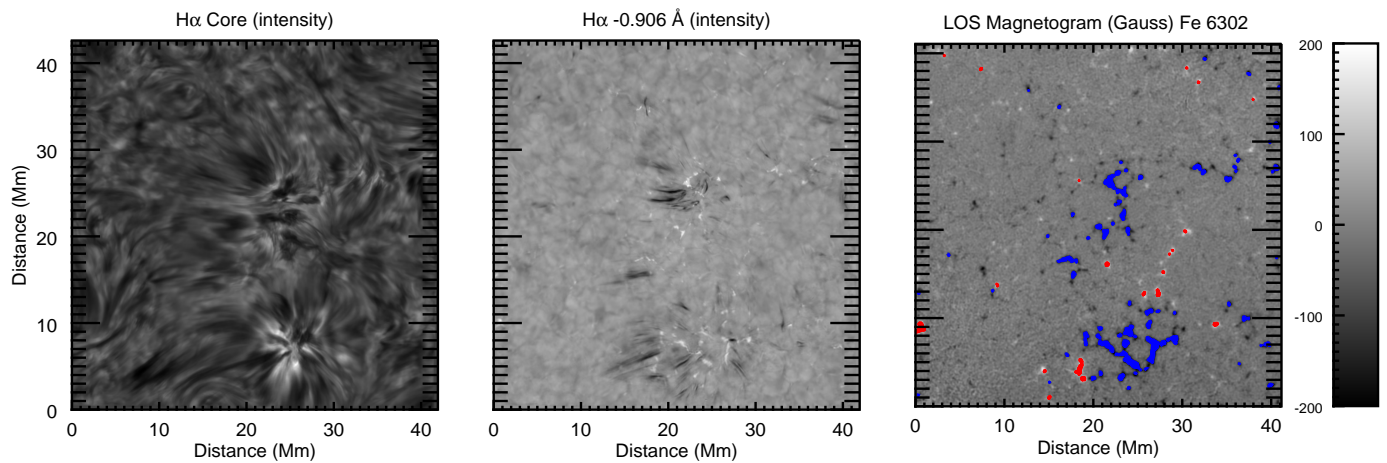} 
        \caption{H$\alpha$ line core and blue wing images are shown in the left and centre panels respectively. The two major magnetic 
        flux concentrations that form part of the network and associated chromospheric rosettes are evident. The right hand panel shows the 
        photospheric magnetogram. The levels have been clipped to $B\sim\pm200$~G to reveal some of the weaker fields in the IN. Contours highlight absolute magnetic field strengths greater than the cut-off, red showing positive flux and blue negative. }
        \label{fig:fov}
\end{figure*}

\section{Observations and data reduction}\label{sec:obs}
The observations presented in the following are focused on the boundary of a coronal hole at the disk centre
(Heliocentric Cartesian - X0", Y31.5"). The dataset used here was taken by the Swedish Solar Telescope (SST - \citealp{SCHetal2003})
 using the CRisp Imaging SpectroPolarimeter (CRISP -\citealp{SCHA2006}, \citealp{SCHetal2008}) 
between 09:06 and 09:35~UT on 2013 May 3 at the Roque de los Muchachos Observatory, La Palma in the Canary Islands. The main 
sequence is a spectral scan of H$\alpha$ (6563~{\AA}) including the following wavelength positions from line centre: (-0.91, 
-0.54, -0.36, 0.0, 0.36, 0.54, 0.91)~{\AA}, corresponding to a range of $\pm$ 41 km/s in velocity. The data was reconstructed with an
extended multi-object multi-frame blind deconvolution (MOMFBD) process (\citealp{VANNetal2005}; \citealp{HEN2013}; 
\citealp{DELetal2015}) and de-rotated, 
aligned, and de-stretched (\citealp{SHIetal1994}). Post-reconstruction, the cadence of the full spectral scan is approximately 
1.3~s with 1150 frames, the spatial sampling is  $\sim$0".06 per pixel, and the resolution is 0''.16. CRISP also undertook 
photospheric Fe I 6301 and 6302 {\AA} spectral scans every approximately five minutes apart over the same field of view
(FOV) to obtain full Stokes profiles. The 
Stokes V component was used to construct line-of-sight (LOS) magnetograms (see \citealp{KURetal2015} for additional details). The scans of the 
iron lines then leave 30~s gaps in the H$\alpha$ time-series. Other works that used this dataset include
\cite{SAMetal2016},
who studied the impact of the lifetime of chromospheric structures on Fourier power spectra, and \cite{HENetal2016},
who found a connection between RBEs and Rapid red-shifted events (RREs) and emission in SDO pass bands sensitive to transition region and 
coronal temperatures. 

The highly dynamic nature of the chromosphere leads to Doppler shifts of line profiles and leads to problems when observing 
phenomena at a fixed wavelength position, mixing fluctuations in plasma conditions and velocities. To overcome this
we also determine the intensity at the location of maximum absorption. Each spectral line profile is fit with a polynomial
to determine the intensity at the minimum location and the wavelength position of the profile minimum.\footnote{No Doppler
shift data is shown here but it was examined as part of the investigation.}

\section{Fibrils in the internetwork}\label{sec:inf_ds}
The observed FOV is displayed in Figure~\ref{fig:fov} and reveals a top-down view of the lower solar atmosphere. 
The Fe magnetograms and wing images from H$\alpha$ show the state of the photosphere,
revealing that there are two dominant regions of negative polarity, kilogauss magnetic fields that 
are likely part of the network. They also reveal the existence of widespread, small-scale patches of magnetic flux away from
the network fields, among which exist a small number of compact magnetic concentrations with $|B|>200$~G. 

The network magnetic fields are visible as clusters of bright points in the 
photosphere and H$\alpha$ line core images suggest that the network has 
a dominant influence on the visual appearance of the chromosphere. In this FOV, the chromospheric fine-scale 
structure predominantly originates
from these network regions. Spicules are observed directly over the network, with an apparent orientation that is 
near perpendicular
to the surface, and long fibrils are also evident around these kG fields, with clear structuring that forms two distinct rosettes. 
The fibrils extend out near-radially from the network to around 10-15~Mm into the IN. Although, this behaviour differs in 
regions where there is a strong positive polarity field (e.g. [18,4] and [26,15]~Mm) and shorter fibrils exist.

\begin{figure*}[!ht]
\centering
        \includegraphics[scale=1., clip=true, viewport=0.cm 0.cm 15.5cm 14.cm]{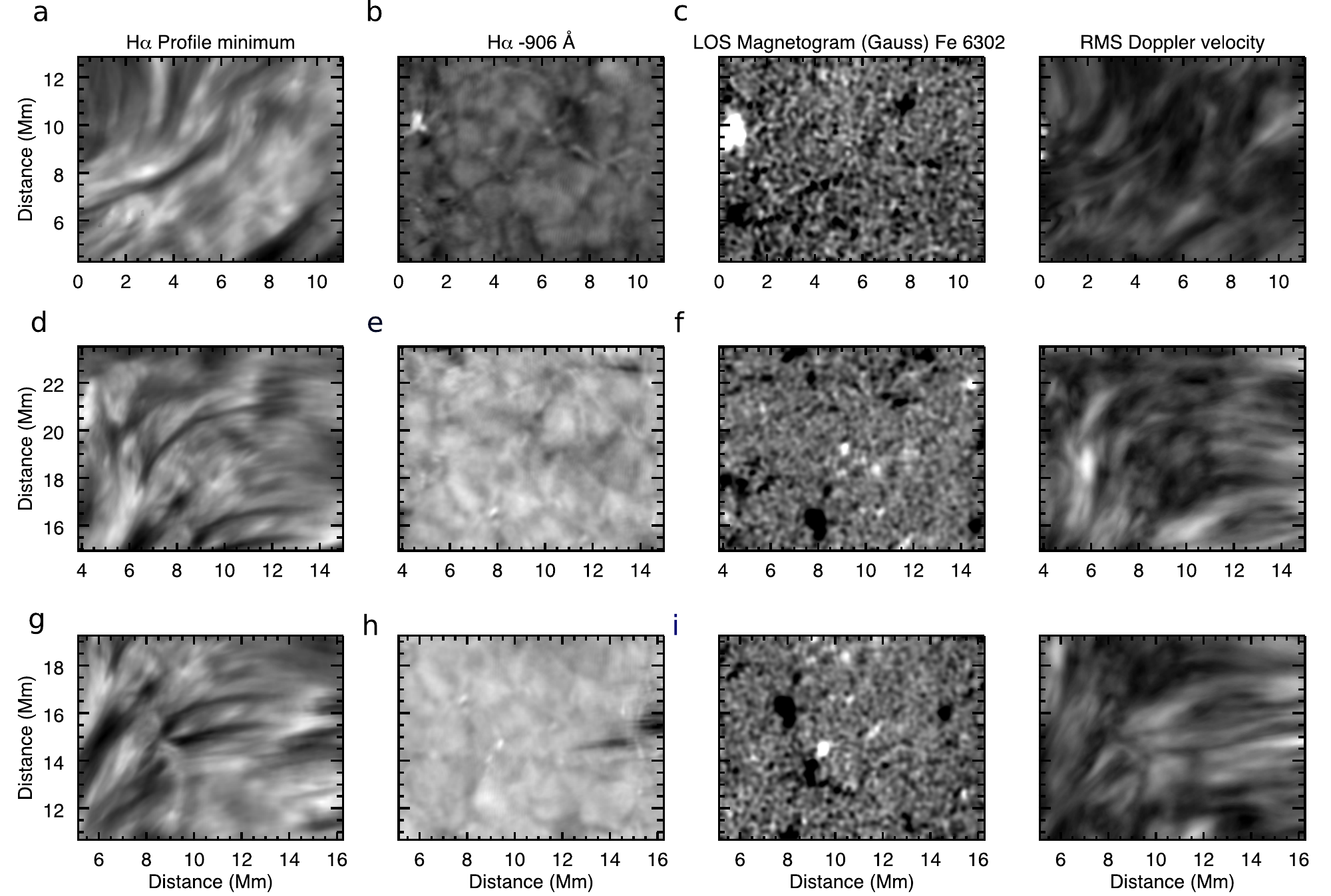} 
        \caption{Examples of internetwork fibrils. The left hand column displays the intensity images determined from the profile 
        minimum, revealing
        the chromospheric absorption features. The middle panels show the corresponding wing images and the last column shows
        the photospheric magnetic field strength clipped between $\pm 50$~G. }
        \label{fig:infib}
\end{figure*}

Additionally, a number of relatively long-lived ($>200$~s), 
extended curvilinear features are found in the internetwork that show absorption in H$\alpha$ line-core images. 
We will refer to these features as IN fibrils and examples are shown in Figure~\ref{fig:infib}. 
The IN fibrils are found to disappear and re-appear in approximately the same 
location over the length of the dataset and follow the same course. This behaviour would imply the persistence 
of an underlying magnetic field and its visibility may be just due to a variation in the H$\alpha$ opacity. 
The existence of long-lived chromospheric magnetic fields with varying opacity, occupying similar
spatial locations, is not at odds with the proper motion speed of photospheric IN magnetic elements (average speeds of
$\sim0.2$~\kms ; \citealp{WANZIR1988}). Their visual appearance and 
behaviour are similar to the fibrils that protrude from the
network, which would suggest that these features are dense plasma that outlines the IN magnetic fields. 
There is also some similarity in behaviour with slender Ca II H fibrils observed in an active region, reported recently from 
the balloon-borne Sunrise data (\citealp{GAFetal2016}). The
elongated nature of the IN features in this FOV would imply, if they do outline the magnetic field, that the magnetic
field is relatively low-lying in the atmosphere. This is opposed to being near-vertical features like spicules, which protrude significantly 
into the corona and predominantly have
limited inclinations from the vertical (\citealp{TSIetal2012}, \citealp{PERetal2012}). {This situation 
represents the view that the H$\alpha$ chromosphere is essentially a corrugated fibrilar canopy, 
with fibrils generated in the 3D simulations found to  
exist up to heights considered to be coronal (\citealp{LEEETAL2012}).  Furthermore,  
H$\alpha$ limb observations (E. Scullion - Private communication) also appear to show 
fibrils rising above the 'bulk chromosphere' 
(i.e. the non-fibrilar chromosphere described in \citealp{JUDCAR2010}).} 
The IN fibrils may be a visible signature of low-lying loops that are suggested to 
dominate the IN by the magnetic field extrapolations of \cite{WIEetal2010}. Alternatively, the IN fibrils' curvilinear nature
could be in the 'plane' of the chromosphere, that is, the 
visible sections of the fibrils are horizontal and snake through the chromosphere.  However, it is near impossible to determine
this from a visual inspection of the current data set. {In Section~\ref{sec:seis}, we employ magneto-seismology
to gain some insight into this aspect of the fibrils.}

%IN Fibrils overview

From examining the data, it proves difficult to determine the photospheric footpoints of the IN fibrils. Both 
in H$\alpha$ wing images, which
show a photospheric scene, and magnetograms, there are no obvious magnetic field concentrations that can be directly associated
with the visible endpoints of the IN fibril, if we compare, for example, absorption features seen in the profile minimum to line wing and 
magnetogram images in Figure~\ref{fig:infib}. Taking the example in the second row, the left hand-end of the IN fibril lies 
close to a negative
small patch of negative polarity (2,2.5)~Mm that is potentially its footpoint. However, there is no clear positive polarity 
patch at the other visible end of the feature, $\sim$(6,8)~Mm. It has been noted previously by \cite{REAetal2011} that
network fibrils typically only have one clearly identifiable footpoint, that being the network patch it originates in, 
while the other footpoints are in the IN with their exact location significantly harder to establish.  

\medskip
A number of {on-disk} chromospheric features located around the network are observed to have a significant blue wing 
enhancement (\citealp{ROUetal2009}) that is thought to be related to an initial heating event 
(\citealp{RUTVAN2016}). A cursory investigation found no obvious heated precursors in the IN 
fibrils that we examine.

\begin{figure*}
        \centering
        \includegraphics[scale=0.95, clip=true, viewport=0.cm 0.0cm 19.9cm 5.cm]{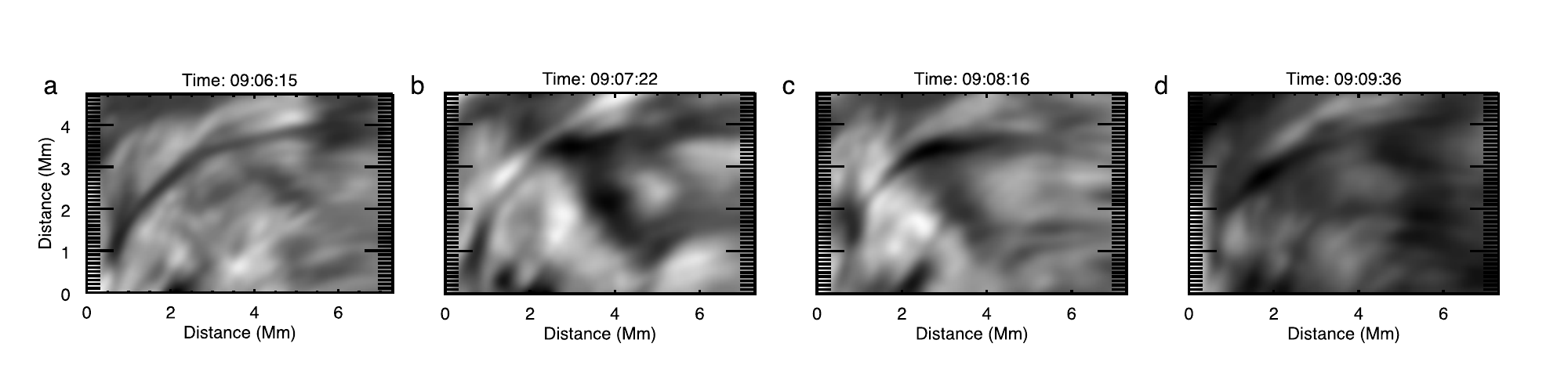} 
        \caption{Panel (a) shows one of IN fibrils in the data set, which can be seen at (8,20)~Mm
                        in the line core image in Fig.~\ref{fig:fov}. In (b), a parcel of plasma is seen at the upper
                        end of the fibril, significantly broadening the apparent width of the feature. The dark clump
                        can be seen to move along the apparent axis of the feature in panels (c) and (d) (from 
                        upper right to lower left). }
        \label{fig:flow}
\end{figure*}

However, we observe an example of absorbing material flowing along the length of one of these IN features after formation 
(Fig.~\ref{fig:flow}). The observation sequence begins at $\sim$09:06 and this particular IN fibril is present then, hence has 
been in existence for an unknown length of time. 
At $\sim$09:07 a broad absorption feature is seen at the upper right of IN fibril and propagates along the axis of the fibril over
the next two minutes. The material does not have a large enough velocity component along the line of sight to show a distinct signature in either the line core Doppler shift or the line wings. It is also difficult to track through individual images, so
no precise estimate of its velocity is possible. A crude estimate puts the apparent velocity somewhere between 10-20~\kms. 
We suggest this observed case is evidence of a flow aligned with the elongated 
axis of the absorption feature, that provides some support for the presence of a guiding magnetic field over an extended
period of time. 

The IN fibrils are also observed to support quasi-periodic transverse displacements and the remainder of this article focuses on them. 
Such motion can (and has previously) been interpreted as MHD wave behaviour, in particular the kink mode, and has
been observed in network H$\alpha$ fibrils (e.g. \citealp{MORetal2012c}) and Calcium active region fibrils 
(\citealp{PIEetal2011}, \citealp{JAFetal2016}).

\section{Data analysis}

\subsection{Estimating the uncertainty} \label{sec:errors}
In order to make reliable measurements of the transverse displacement of fibrils and the associated errors, 
an estimate of the measurement errors associated with the data is required. The SST and CRISP provide high 
signal-to-noise observations with low photon noise. However, there are a number of other sources of noise 
associated with the data. 

Distortion due to atmospheric seeing is a major source
of noise. While the extended MOMFBD procedure and de-stretching of the data make a significant impact on 
reducing this source of data noise, 
uncertainty will evidently remain on both the intensity values and the physical location of features
of interest. Ideally, some form of statistical re-sampling method would be employed, modelling the noise 
associated with the pre-processed data and re-running the data reduction pipeline numerous times in order to see how those 
uncertainties are propagated through the pipeline and impact upon the relative variations in spatial location and intensity. 
However, this is currently impractical due to the amount of processing time required for each data set.

CRISP is a double Fabry P\'{e}rot Instrument (FPI) set up to finely tune its spectral observations, allowing for high
resolution spectral imaging. However, imperfections on the surface of the etalons of the FPI
result in ``cavity'' errors (\citealp{DELetal2015}), leading to field-dependent shifts in the central wavelength of the 
transmission profile. This leads to uncertainties in measured intensity at high-gradient spectral regions such as the wings of the 
line, which are corrected intensity-wise by the pipeline and wavelength-wise on a per pixel basis when computing Doppler maps (or 
other situations where high accuracy in the wavelength is necessary) by applying the pipeline calculated cavity shifts. Such 
uncertainties should be extremely small at the core of the line where the gradient is low, but will nonetheless contribute.

In order to
gain an estimate for the uncertainty associated with the data, we turn to multi-scale image processing techniques. 
We begin by assuming that we can estimate the 
uncertainty on the measured intensity values from the noise within the 
processed data and we would like to establish
a relationship between the intensity and the noise. To do this we first create a `noise image' from the whole FOV of the data. The data is cropped to remove artefacts at the edges left over from the reduction pipeline. 

\begin{figure*}[!ht]
\centering
        \includegraphics[scale=0.85, clip=true, viewport=0.cm 0.cm 19.5cm 17.cm]{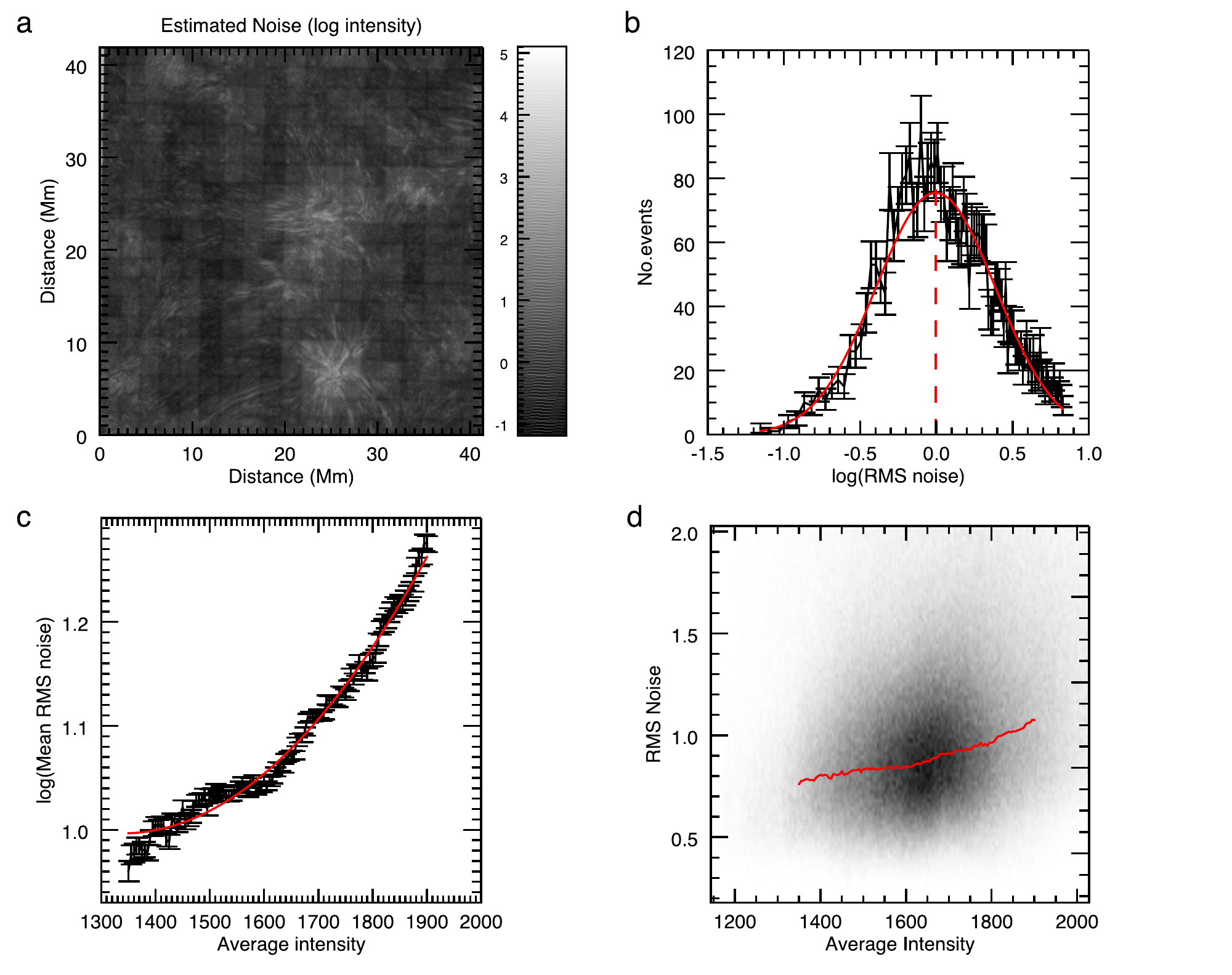} 
        \caption{Noise estimation process. The RMS residual intensity or noise (natural log units) is obtained across the 
        FOV (a), which has the joint probability distribution shown in (d) when compared to average intensity. The bin widths
        used in (d) are 0.01 units in the RMS noise and five units in the intensity. In (b), an example of the 1D 
        histograms of the RMS noise estimate for the intensity bin 400-405 is shown. The over-plotted red curve in (b) 
        is a fitted Gaussian and the vertical red dotted line signifies the location of the mean. Panel (c) demonstrates the 
        relationship between average intensity and the mean value of the log RMS noise estimate. The polynomial fit to the mean 
        values is over-plotted in red, the equation of this curve (Eq.~\ref{eq:ierr}) will serve as the estimate of noise. The 
        same data points are over-plotted on the JPDF (d) as a red line.}
        \label{fig:dens}
\end{figure*}

The data is then filtered in space by applying a multi-scale filter utilising the \`{a}-trous algorithm with a 2D B3 spline filter 
(e.g. \citealp{STECOB2003}, \citealp{STAMUR2006}), and then we apply
unsharp masking to the highest-frequency scale. This process ensures the data is reduced to variations on the smallest spatial scale and 
the residuals can be taken to be indicative of the data noise (e.g. \citealp{OLS1993}). 
The root mean square (RMS)
of forty successive noise frames in time is then taken to estimate the standard deviation per pixel. Figure~\ref{fig:dens}a
shows the estimated RMS noise for this data set. There is clear evidence of fixed pattern noise in the 
RMS noise image. It is revealed as a grid pattern and is likely due to small differences between the different MOMFBD sub-fields. 
The presence of the grid pattern in this RMS image is likely exacerbated due to
one or two frames taken during poor seeing conditions, which have relatively worse resolution and a somewhat smoother
intensity profile. Hence, there is less intensity variation across
these regions and a reduction of the average noise in the regions. The grid pattern only occupies a 
minority of the pixels across the image and will likely only impact minimally on the following analysis of the data noise.

Figure~\ref{fig:dens}d displays the joint 
probability distribution function (JPDF) of the average intensity (averaged over the same 40 frames) and the RMS noise. The
JPDF shows a discernible trend of increasing noise with intensity. In order to characterise this trend, the RMS noise values are binned 
as a function of average intensity, and the 1D PDFs of RMS noise show an approximate log-normal behaviour 
(e.g. Fig.~\ref{fig:dens}b). The values 
of the mean and standard deviation of the log-normal RMS noise are determined for each average intensity bin. 
Figure~\ref{fig:dens}c shows the average
intensity versus the mean values of RMS noise and a quadratic function is fit to the mean values to establish a 
relationship between the two quantities. The found parameters give:

\begin{equation}\label{eq:ierr}
\ln{\sigma_I}=8.2\times10^{-7}I^2+5.5\times$10$^{-4}I+ 0.087.
\end{equation}

\noindent This equation will be used to provide our estimates of uncertainty associated with measured intensity. The resulting
uncertainty is approximately 0.1$\%$ of the intensity, which is in line with expected noise levels estimated from
theoretical arguments by \cite{DELetal2012} for CRISP-like FPIs operating on similar optical systems to the SST.  

%Error figure

\subsection{Transverse motion} \label{sec:wave}
\begin{figure}[!ht]
    \centering
        \includegraphics[scale=0.5, clip=true, viewport=0.5cm 0cm 18.5cm 16.6cm]{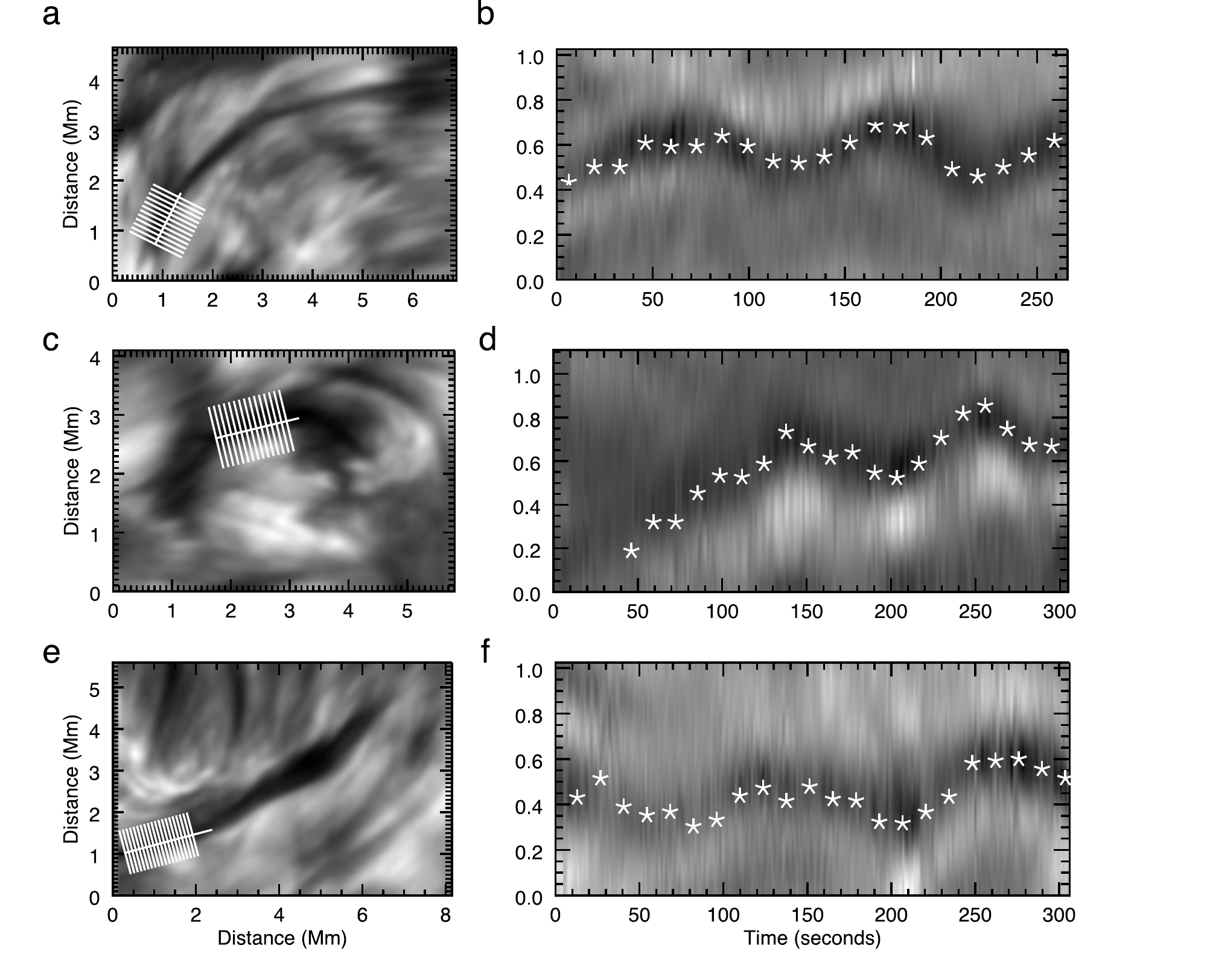} 
        \caption{Examples of IN fibrils are shown in the left-hand column. White lines are over-plotted to show locations of 
        cross-cuts taken along the structure. The longitudinal line serves as a guide line along the fibril axis. 
        The right-hand panels show 
        examples of time-distance diagrams from the IN fibrils, revealing the transverse displacements of the IN fibril. The
        white stars highlight where the measured central locations of the IN fibrils axis are from the fitting routine. Every 10th point
        is plotted for clarity.}
        \label{fig:cross}
\end{figure}

The IN fibrils are observed to undergo a quasi-periodic displacement transverse to the elongated direction. This motion
is revealed in time-distance diagrams created from slits placed perpendicular to the observed axis of the elongated direction
(Fig~\ref{fig:cross}).

The fibrils' cross-sectional flux profile is approximately an inverse Gaussian, and the central location of the 
fibrils axis is defined as the location of greatest absorption, that is, the minimum of the flux profile. Hence, from the time-distance 
diagrams, the central locations of the fibril axis, $x(t_i)=x_i$, were found using the NUWT (Northumbria University 
Wave Tracking) code (\citealp{MOR_NUWT}, further details are given \citealp{MORetal2012c,MOR2014}). 
NUWT requires the uncertainties in intensity (Eq.~\ref{eq:ierr}) to establish the central location with sub-pixel accuracy and
also provide meaningful uncertainties on this position ($\sigma_i$). 
Examples of the IN fibrils and the resulting tracked axis in the time-distance diagrams are displayed in Fig.~\ref{fig:cross},
with the $x_i$'s forming a time-series of the transverse displacement. 
Time-series are obtained in this manner for multiple cross-cuts placed at equally spaced intervals along the axis of the 
fibril, enabling a picture to be built up of how the wave evolves as it propagates.

In order to characterise the apparent periodic nature of the displacements, the individual displacement time-series 
($(t_i,x_i)$, $i=0,...,n$) 
are fit by least-squares (\citealp{MAR2009}), using a model of the form,
\begin{equation}
\hat{x}(t_i)=a_{1} + a_{2}t_i+a_{3}sin(2\pi/a_{4}t_i-a_{5}), 
\end{equation} \label{eq:sin}
\noindent where $a_3$ is the amplitude, $a_4$ is the period, and $a_5$ is the phase. The linear terms represent any potential 
long-term drift of the fibril over the course of the quasi-periodic motion.

%Fibril cross-cut example{}

\subsection{Propagation speed} \label{sec:phase speed}

To measure the propagation speed of the transverse waves supported by the fibril, the time-series obtained at 
multiple locations along the fibril 
are cross-correlated. The time-series from the central cross-cut is used as a reference series and 
the preceding and proceeding series are cross-correlated with the reference, enabling a measurement of 
the time-delay between time-series. We note
this procedure assumes that there is no significant curvature of the fibril perpendicular to our view of it. Any such curvature would
lead to an underestimate of the distance that the wave propagates, leading to slower measured propagation speeds.

\begin{figure}
        \centering
        \includegraphics[scale=0.57, clip=true, viewport=2.5cm 7.8cm 18.5cm 19.6cm]{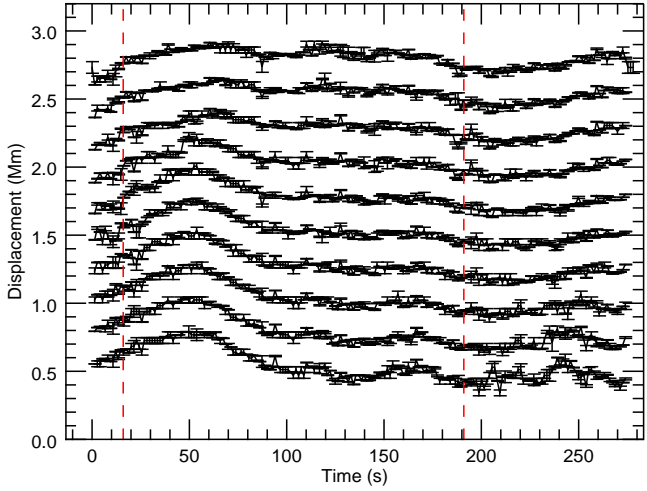} 
        \caption{ Set of time-series taken from a single IN fibril. Each time-series corresponds to 
        a measurement of the central location of the fibril from a time-distance diagram. 
        In this example, the time-series are measured at separations of 85~km taken 
        along an IN fibril feature. 
        The red dotted lines shows the time interval used for cross correlation.}
        \label{fig:ts}
\end{figure}

Shorter sections of the time-series are chosen for cross-correlation due to the
possibility that neighbouring peaks in displacement are related to counter-propagating waves (e.g. \citealp{MORetal2012c}). 
Additionally, upon examining the displacement time-series, there are data points 
that show relatively large variations in position from the local 'average'. This feature is observed across all time-series for a 
particular feature and occurs when the data quality
drops in that particular region, which we associate with seeing variations across the image. The prominent nature of these 
excursions can impact upon the cross-correlation
results. Hence, a cautious approach is applied to selecting segments of the time-series to cross-correlate, avoiding these 
outliers. This is currently
a rather subjective process and future studies will require more advanced objective methodology to counter such issues, for example,
wavelet or empirical mode decomposition may provide this ability, or an initial cleaning of the data before feature measurement
occurs. Once a suitable portion of the time-series has been identified, the series are cross-correlated and a polynomial function is fitted
around the local maximum of the correlation function to achieve sub-cadence accuracy on the lag value.
Figure~\ref{fig:ts} shows an example of the different time-series taken along a fibril used for cross-correlation. 

In an attempt to determine the uncertainties on the lag values, a parametric re-sampling technique is employed.
In brief, the idea behind the re-sampling methodology is to try and determine the distribution of an 
estimator in order to establish indicators of the estimators reliability, for example standard errors or confidence 
intervals. The parametric re-sampling makes use of an estimated or hypothesised model for the uncertainties in 
the original data, and propagates these errors via Monte Carlo simulation. 
This methodology eliminates the need for simplifying assumptions, which are required to 
obtain approximate analytic formulae
describing how uncertainties propagate through complicated operations on the data.
Here the estimator of interest is the lag value, and we would like to know how the uncertainties
associated with the measured locations of the IN fibrils' central axis influence the lag value. 
This methodology enables us to incorporate the heteroscedastic nature of the measurement uncertainties, $\sigma_i$,
which we assume are normally distributed about the measured location, 
$\mathcal{N}(x_i,\sigma_i)$. The methodology is then to generate repetitions of the displacement time-series 
with different values of measurement noise and to cross-correlate these re-sampled series. 

\begin{figure}[!ht]
        \centering{
        \includegraphics[scale=0.48, clip=true, viewport=2.5cm 7cm 19.5cm 20.6cm]{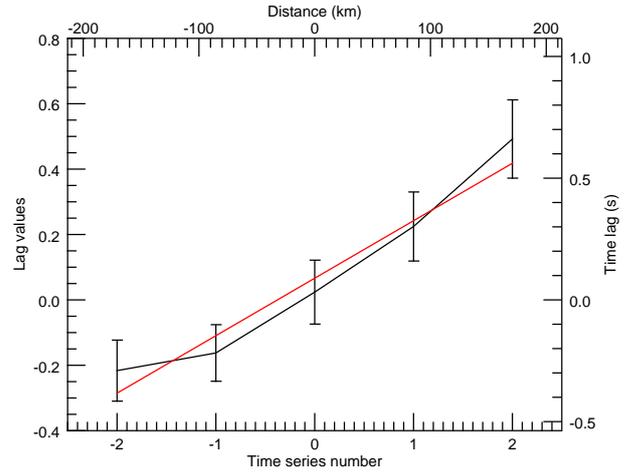}
        \includegraphics[scale=0.49, clip=true, viewport=2.5cm 7.5cm 20.5cm 19.6cm]{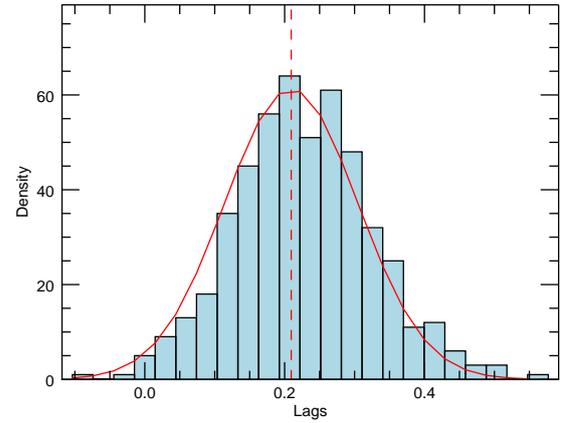}
        }
        \caption{Propagation speed measurement. The top panel displays the lag values as a function of 
        the time-series position. The zero time-series number corresponds to the middle time-series in the range. 
        A linear fit is shown by the red line, the gradient of which determines the propagation speed. The
        bottom panel shows at typical distribution of time lags between two time-series 
        from the re-sampling technique. The red line is a fitted Gaussian with the mean shown by the dashed red line. 
        The mean and sigma values of the Gaussian are used as the lag and uncertainty on the 
        lag respectively.}
        \label{fig:boot}
\end{figure}

\noindent The following 
steps describe this implementation. 
\begin{enumerate}
\item Generate random noise for each displacement time-series data point $(t_i,x_i)$ from a 
normal distribution, $\mathcal{N}(0,\sigma_i)$. 
\item Add this random noise to the original time-series, creating a re-sampled time-series with a different realisation of the data noise.  
\item Generate five hundred replicates in this manner for each time-series. 
\item Cross-correlate the five hundred reference signal replicates with its counterpart from different spatial locations. 
\item Establish a distribution of time-lags from the five hundred correlations, calculating the mean lag value and its standard deviation 
(e.g. Fig.~\ref{fig:boot}). 
\end{enumerate}

A linear function is then fit to the lag values as a function of distance along the fibril, with the gradient providing 
a measure of the propagation speed (Fig.~\ref{fig:boot}). It is found that the lag values show little evidence of
deviations from a straight line, with the linear fits having a reduced $\chi^2\sim1$. In certain cases, the $\chi^2$ is
substantially less than one, potentially implying that the associated uncertainties with the lag values are over-estimated. 
However, at present we remain on the side of caution with our interpretation, and the constant time-lag with
height implies there is little variation of propagation speed as a function of distance along the IN fibrils.

\subsection{Magneto-seismology} \label{sec:seis}
{In our visible inspection of the data in Section~\ref{sec:inf_ds}, it was noted that the relative inclination
of the IN fibrils within the atmosphere was near impossible to establish from visual examination alone. We now
outline the magneto-seismological methodology required to estimate variations of plasma parameters along the 
fibrils, which may provide information on whether the fibrils are lying horizontally in the chromosphere or not.}

There is a substantial volume of literature dedicated to the use of kink waves as an inversion tool to diagnose the 
local plasma properties. In general, the theoretical development of the atmospheric magneto-seismology has contained very strict assumptions 
about the nature of the plasma (e.g. that it is hydrostatic), and the wave evolution (e.g. that there is no damping during propagation). There are
attempts to relax some of these assumptions, for example \cite{MORERD2009b}, \cite{RUD2011}, and \cite{SOLetal2011b}\footnote{ 
The emphasis of these works is on dynamic coronal loops, but the intent is the same.}, but this still does not capture the complex 
physical effects observed during the evolution of the kink waves along chromospheric wave-guides (e.g. \citealp{MOR2014}). 
Despite these constraints, it is useful to compare observed wave behaviour to the current theory in order to aid interpretation. 

%Time-series example

\cite{VERTetal2011} demonstrated that inversion of the properties (amplitude, phase speed) 
of a propagating kink wave along a magnetised wave-guide can inform us about 
the variation of the local plasma density and magnetic field strength. We recall that the WKB solution to the wave 
equation governing a kink wave propagating along a stratified flux tube under the thin flux tube approximation
(\citealp{MORetal2012}) is given as\footnote{The form of this solution requires the longitudinal changes in magnetic 
field and plasma density along the flux tube to be small compared to the wavelength. This assumption 
may be violated in chromospheric
waveguides.},
\begin{equation}
\xi(\textit{z})=C\sqrt{\frac{c_k(\textit{z})}{\omega}}R(\textit{z}).
\label{eq:amp}
\end{equation}
Here $\xi(\textit{z})$ is the displacement amplitude as a function of distance $z$, $c_k(\textit{z})$ is 
the kink phase speed, $\omega$ is the 
angular frequency, $R(\textit{z})$ is the radius as of the flux tube as a function of height, and C is a constant.
Starting from 
Eq.~(\ref{eq:amp}), we derive relations for the normalised density, radius, and magnetic field
following \cite{MOR2014}, which can give an indication of the quantities' relative evolution of the stratified flux tube,

\begin{align}
\frac{\langle\rho(z)\rangle}{\langle\rho(0)\rangle}&=\frac{\xi(0)^4}{\xi(z)^4}, \label{eq:rho} \\
\frac{R(z)}{R(0)} &= \sqrt{\frac{c_k(0)}{c_k(z)}}\frac{\xi(z)}{\xi(0)}, \label{eq:radius}\\
\frac{\langle B(z) \rangle}{\langle B(0) \rangle} &= \frac{c_k(z)}{c_k(0)}\frac{\xi(0)^2}{\xi(z)^2}.
\label{eq:mag}
\end{align} 
Here, $\langle\rho(\textit{z})\rangle$ is the density averaged over the internal and ambient plasma and 
$\langle B(\textit{z}) \rangle$ is the local average magnetic field. These 
relations show that the above quantities rely solely on the amplitude ($\xi$) and kink phase speed ($c_k$). 
We emphasise that these relations only hold under an assumption of no wave damping and a hydrostatic plasma. As mentioned 
in Sec.~\ref{sec:phase speed}, a linear function provides a reasonable fit to the slope of the time-lags, with 
no evidence that a higher order polynomial is required (within the current uncertainty estimates). 
As such, the propagation speed of 
the transverse wave can be considered constant along the wave-guide, which 
means the propagation speed terms in Eqs.~(\ref{eq:rho})-(\ref{eq:mag}) drop out, leaving relations entirely dependent on amplitude. 

%Phase speed measurement figure

Finally, depending on the nature of the IN fibrils, the variation in magnetic field and density will be 
dependent upon the inclination of the fibril to the vertical (Section~\ref{sec:inf_ds}). In a hydrostatic atmosphere, a straight flux 
tube will have an exponential decrease in density, modified by any inclination. However, should the IN fibril
be loop-like, the expected variation in density in a hydrostatic atmosphere is given by
\begin{equation}
\rho(z)=\rho_0\exp\left(-\frac{L}{\pi H}\cos\left(\frac{\pi z}{L}\right)\right),
\label{eq:dens_prof}
\end{equation}
assuming the loop is semi-circular (e.g. \citealp{DYMRUD2006a}), where $z$ is the distance along the loop,
$H$ is the scale height, and $L$ is the loop length. Further modifications to this relationship
are present should the loop geometry deviate from this (e.g. be elliptical; \citealp{MORERD2009}). 
These gravitationally stratified profiles for the variation in density will lead to upward propagating 
waves showing amplification and downward propagating waves displaying attenuation (Eq.~\ref{eq:rho}).

One must also recognise that any estimated density profile will be an average of internal fibril and ambient plasma densities, 
hence, variations of the ambient plasma with height
can also contribute to the measured gradients (\citealp{MOR2014}). If, for example, we were to assume that the ambient atmosphere
around the IN fibrils takes on a profile of an approximately gravitationally stratified atmosphere, 
and the IN fibrils are features composed
of non-hydrostatic chromospheric plasma that protrude
into the corona (although not as far as spicules), then the external density may drop off rapidly 
compared to the internal density, leading to
significant changes in wave amplitude.

%Wave measurement results Figure 
\begin{figure*}
        \centering
        {\includegraphics[scale=0.9, clip=true, viewport=1.5cm 6.8cm 19.5cm 21.6cm]{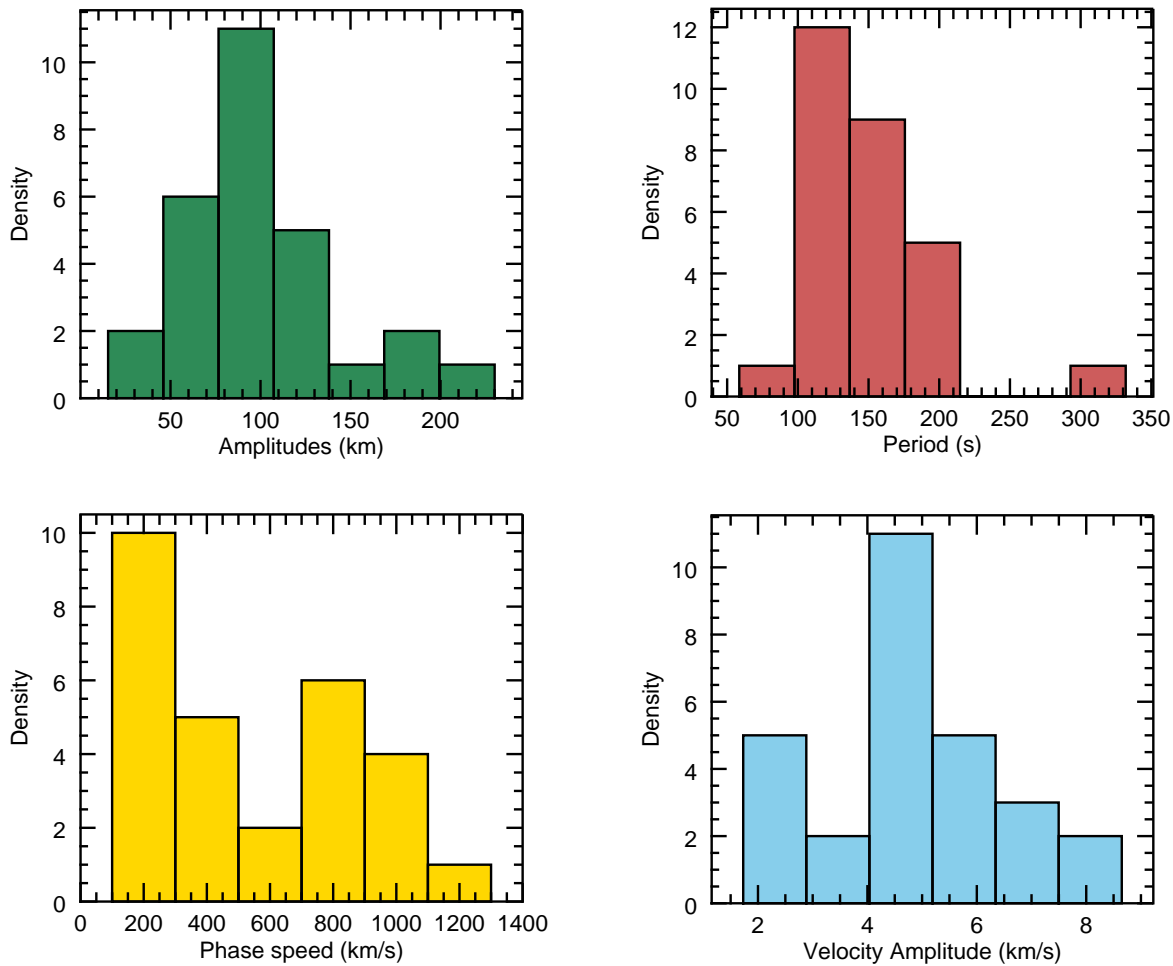}}
        \caption{Histograms displaying the measured amplitudes, periods, phase speeds, and velocity amplitudes from the 28 IN 
                        fibrils measured.}
        \label{fig:hist}
\end{figure*}

\section{Results} \label{sec:RES}
In this section we present the results from the measurements of the MHD kink waves observed in 28 IN fibrils.
An overview of the measured wave properties is presented in the histograms in Fig.~\ref{fig:hist}, showing the
distribution of displacement, period, velocity amplitude, and propagation speed. The values shown in the histograms
are the weighted means of the measured properties along fibrils, providing in essence a summary of the wave properties.
The velocity amplitude is calculated utilising the standard equation,
\begin{equation}
v= \frac{2\pi\xi_{wm}}{P_{wm}},
\end{equation}
where $v$ is the velocity amplitude, and $\xi_{wm}$ and $P_{wm}$ are the weighted mean displacement amplitude and period of each feature, respectively. The mean, median, and standard deviation of these results are given in
Table~\ref{tab:prop}. The results are consistent with previous quiet sun measurements of network 
fibrils (e.g. Table 3 in \citealp{JESetal2015}). {We note that the gaps in the data may have an impact on our 
ability to measure waves with periods on the order of, and longer than, 300~s, although \cite{MORetal2013b} 
did not find many examples of such periods in both quiet and active region fibrils. } \\

\begin{table}
\centering
\caption{\label{tab:prop}Average wave properties.}
\begin{tabular}{lccc}
\hline\hline
  & Mean  & Median & Standard\\ 
  &   &   & deviation \\ 
  \hline
$\xi$~(km) & 85  & 78 & 43 \\
$P$~(s) & 128  & 122 & 43 \\
$v$~(\kms) & 4.22  & 4.21 & 1.6 \\
Propagation speed (\kms) & 446  & 399 & 338 \\
\hline
\end{tabular}

\end{table}

The results for the propagation speeds of kink waves along the IN fibrils suggest that, in general, they are comparable
to the limited measurements of propagation speeds along network fibrils and mottles (\citealp{JESetal2015}). The 
results are notably greater than those measured for Ca II slender fibrils in active regions by \cite{JAFetal2016}. 
Moreover, there are a number of IN fibrils with propagation speeds that exceed 500~\kms, which are perhaps 
unexpectedly large. This is discussed further in the final section.

%Seismology Figure 
\begin{figure*}
        \centering 
        {\includegraphics[scale=0.98, clip=true, viewport=1.cm 9.8cm 20.5cm 18.cm]{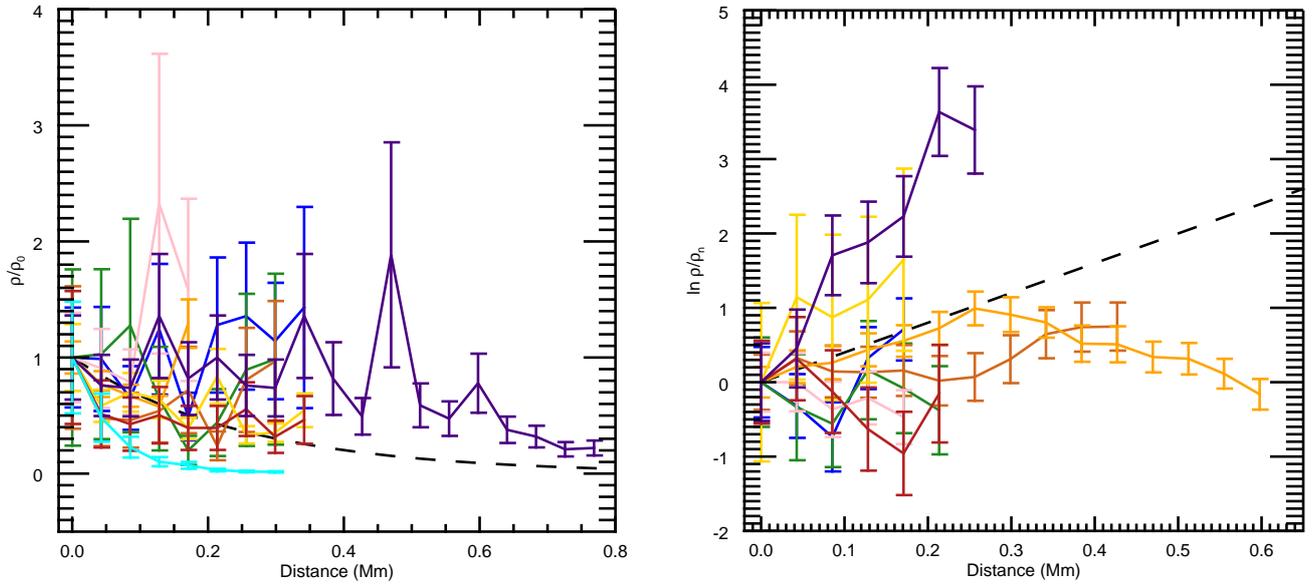}}
        \caption{Magneto-seismological inversions for the relative change in density along the IN fibrils. The profiles are 
        separated whether the wave was measured to be propagating away from the apparent endpoint of the fibril (left) or
        towards it (right). The black dashed line is the
        expected relative change in density for a hydrostatic atmosphere with scale height 250~km. }
        \label{fig:seisprof}
\end{figure*}

\medskip

The magneto-seismology inversions for the density profiles are shown in Fig~\ref{fig:seisprof}. {To
interpret the results, we work under the assumption that the fibrils are closed magnetic field lines, that is low lying loops,
in line with our expectations from the visual inspection of the data in Section~\ref{sec:inf_ds}.} The results are 
separated into two plots dependent on whether the wave is observed to propagate away from the observable 
endpoint of IN fibril (`upward' - left panel) or towards it (`downward' -  right). To aid comparison of
density changes along the fibrils, an exponential density profile for a hydrostatic atmosphere is over plotted, 
with a scale height of 250~km (\citealp{UIT2006}). This represents the maximum change in density through the atmosphere 
for this value of scale height, with inclined and loop-like features having less variation as inclination to the vertical 
increases (Eq.~\ref{eq:dens_prof}). The magnetic field
variation can be inferred from these plots as it is proportional to the square root of the density variation (Eqs.~\ref{eq:rho} \& 
\ref{eq:mag}).

In general, the measurements reveal evidence for trends in the wave amplitude as they propagate 
along the IN fibrils, although the measurements have a high variability (with correspondingly sizeable error bars). The variation
of amplitude suggests that the fibrils and ambient plasma possess some stratification of density and magnetic field, 
indicating that the IN fibrils are not merely horizontal features in the chromosphere. In both the `upward' and 
`downward' propagating waves, there are examples whose variation in amplitude is roughly in line with that expected from a
hydrostatic density profile. 

The blue and pink `upward' profiles would suggest the density increases along those fibrils. 
However, this is potentially a signature of wave damping, which effectively works against the density to attenuate any 
amplitude variation for waves propagating upwards along the wave-guide (the interplay of longitudinal 
inhomogeneities and wave damping was discussed in \citealp{MOR2014}). There is no
evident reason why all observed waves should not also be damped to some degree during propagation, 
with the inhomogeneous transverse structuring 
of the local plasma enabling resonant absorption (e.g. \citealp{TERetal2010c}) or the driving of the Kelvin-Helmholtz
instability (e.g. \citealp{ANTetal2015}) to act upon the observed waves. 
For the upwardly propagating waves, this would imply the average density of the fibrils and external atmosphere are
more stratified than suggested by the majority of the magneto-seismology profiles in Fig.~\ref{fig:seisprof}. 
The potential signature of damping is also present in the `downward' propagating waves, with the damping 
potentially enhancing the effect of increasing density to attenuate the amplitude, as seen with the 
yellow and purple profiles.  

Finally, there is potential evidence for wave amplification, the `upward' cyan profile 
(with an alternative explanation as a highly stratified fibril) or the 
gold `downward' profile. It is not known to us what may cause an additional amplification of the 
wave as it propagates, although a scenario can be imagined involving a suitable combination
of internal and ambient density profiles, for example \cite{MOR2014} reported rapid amplification of a 
propagating kink wave along a spicule and demonstrated the required density profiles that 
could recreate this behaviour.

The vast range of possible scenarios makes it difficult, at present, to extract actual 
gradients of the density along the observed waveguides. It is likely that an advanced inversion scheme is
required to exploit the observed propagating kink waves, potentially utilising the power of Bayesian 
analysis and requiring additional information about the plasma. In the hope that such a scheme is developed,
the trends have been fitted with an exponential profile in order to provide a measure
of the variation (results given in Table~\ref{tab:one}). While we refer to this 
measure as a scale-height, we emphasise that the value will incorporate
the influence of internal and external variations in density, as well as wave damping (or even amplification).

\section{Discussion and conclusion}
Here we have begun to examine the properties and dynamic behaviour of internetwork fibrils observed in H$\alpha$,
with the long term goal of trying to understand their formation and the antecedent energy deposition. The relatively
isolated nature of these features, as opposed to the numerous network fibrils that occupy the rosettes, 
may mean that identifying the mechanism(s) responsible for depositing energy is a simpler task. {Before concluding, we highlight that the number of IN fibrils 
in this study is small and urge caution in assuming that these properties are representative of all IN fibrils.}

In general, we find that the IN fibrils display similar behaviour to network fibrils in H$\alpha$. The fibrils are
found to appear, disappear, and re-appear over periods of five to ten minutes suggesting regular but not continuous
depositions of energy. The reason for their pattern of visibility in H$\alpha$ is unclear at present, although it may
be the result of the fibrils being in a near constant state of thermal non-equilibrium, 
undergoing cycles of heating and cooling (such
cycles have been extensively studied to explain coronal loop behaviour, e.g. \citealp{KLIetal2010}). 
The following is a crude description of how such a cycle could impact on visibility. 
An initial energy deposition occurs along the magnetic fields, 
heating the local plasma, potentially - and eventually - leading H$\alpha$ visibility through delayed post Saha-Boltzmann extinction 
(\citealp{RUT2016,RUTVAN2016}). The heating could drive up-flows of denser material from lower atmospheric heights, which would also
lead to enhanced H$\alpha$ extinction (\citealp{LEEETAL2012}). The fibril material then cools and condenses, evacuating the magnetic flux tube; both cooling and density decrease leading to less H$\alpha$ opacity. Another heating event occurs and the cycle repeats. For
coronal loop modelling at least, this behaviour leads to over-dense loops, hence, larger than expected scale-heights. 

Such a heating scenario may require the existence of a semi-stable chromospheric magnetic flux tube that supports
denser plasma than its ambient environment, which may not be guaranteed on theoretical grounds in 
the partially ionised chromosphere (e.g. \citealp{MARetal2016}).
However, we have observed flows of dense (i.e. absorbing) material along the IN fibril sometime after the initial formation,
which would indicate that a guide for the plasma exists. Moreover, we observe signatures of quasi-periodic, 
propagating transverse displacements of IN fibrils, which we interpret as MHD 
kink waves. The presence of kink waves requires an over-dense magnetic flux tube to be present, with the field 
orientation perpendicular to the direction of displacement. The mere fact fibrils are visible as H$\alpha$ 
absorption features could indicate they are denser than their surroundings, with their appearance
due to increased column mass along the line-of-sight that leads to a greater than average formation height 
(\citealp{LEEETAL2012}). The presence of
the propagating kink modes along the elongated axis supports the idea that the observed absorption features 
outline magnetic fields. 

At present there is some discussion of whether the observed chromospheric features are tracers of the magnetic field. 
Early work by \cite{DELSOC2011} using Ca II 8542~{\AA} spectropolarimteric observations suggested a certain degree of 
misalignment between fibrils and magnetic fields. Recently, more robust analysis implies the misalignment 
degree is much less than previously suggested, although the dispersion of misalignment angle increases in less magnetised
regions (\citealp{ASEetal2017}). Moreover, observations with Helium 10830~{\AA} find little evidence for 
misalignment (\citealp{SCHetal2013}).

On the other hand, \cite{MARetal2016} demonstrated that 2D simulations of partially ionised 
chromospheric plasmas including the effect of ambipolar diffusion also show misalignment between the magnetic
fields and temperature (density) structures. 
However, if these features undergo significant heating during their formation, as has been suggested for some spicules and RBEs 
(\citealp{PERetal2014}; \citealp{SKOetal2015}; \citealp{ROUetal2015}) and also long fibrils (\citealp{RUTVAN2016}),
then it appears crucial for models to require the inclusion of a sluggish, non-equilibrium hydrogen ionisation and to take 
into account the history of heating events in the chromosphere (R. Rutten - Private Communication).\footnote{ 
Detailed treatments of  
non-equilibrium Hydrogen for model atmospheres can be found in \cite{CARSTE2002}; \cite{LEEWED2006}; \cite{LEEetal2007}.
The consequences of heating events on the line formation of H$\alpha$ is discussed at length in \cite{RUT2016} 
and \cite{RUTVAN2016}.} This would then naturally lead to a significantly greater fraction of ions and 
electron densities in the post-heating plasma (e.g. see post (inter)shock regions in \citealp{LEEetal2007}), 
and a potential reduction in the influence of ion-neutral related effects. Additionally, 2D simulations are 
unlikely to capture the necessary physics for the heating of chromospheric phenomena, missing torsional motions that can
input additional energy and momentum into the chromospheric plasma (\citealp{MATSHI2010}; \citealp{IIJ2016}).
A recent investigation of the Bifrost simulation by \cite{LEEetal2015} suggests that their model chromosphere shows
a mixed picture of alignment.

\medskip 

Returning to the measurements made here, we have been able to measure the propagation speed of a number of 
kink waves along the fibrils. Somewhat surprisingly, we find no
evidence for variation in the propagation speed as a function of distance. The propagation speed measurements
allow us to place rather coarse constraints on the local Alfv\'en speeds, hence, magnetic field values associated with
the fibrils in the upper chromosphere. The following equation
describes the relationship between the propagation speed (kink speed), plasma quantities, and the Alfv\'en speed,
\begin{equation}
c_k^2=\frac{2B^2}{\mu_0(\rho_i+\rho_e)}=v^2_A\left(\frac{2}{1+\zeta}\right),    
\end{equation}
where, $\mu_0$ is the magnetic permeability of free space, $\zeta=\rho_e/\rho_i$, and the subscripts $i, e$ refer to internal 
and external plasma quantities. Here we have assumed $B_i\approx B_e$. For estimated
chromospheric densities of $10^{-9}-10^{-10}$~kg~m$^{-3}$ along absorbing H$\alpha$ features (taken from
simulations of a model quiescent chromosphere; \citealp{LEEETAL2012}), this
would imply chromospheric magnetic field strengths somewhere in the range of $3-200$~G. 
The larger values of propagation speeds (\textgreater500~\kms) are significantly greater than
previous measurements, and would correspond to magnetic fields of $B\sim25-200$~G. 
While the lower end of this range seems feasible, it is unclear 
whether the larger values of field strength would exist in the IN chromosphere. These larger values are certainly at 
odds with the LOS field strengths observed in and around the IN fibrils' apparent footpoints. However,
we note that no correction for stray light has been attempted on the magnetograms, with stray light potentially 
leading to an underestimation of the field strength up to as much as half for isolated magnetic elements, due to 
the contribution to the Stokes V profiles of surrounding non-vertically-magnetised regions (e.g. the Milne-Eddington 
inversions of similarly processed data by \citealp{NAR2011}). Furthermore, the larger values given here are in 
excess of estimated magnetic field strengths for spicules at the limb (e.g. $\sim50$~G \citealp{CENetal2010}), 
which are generally associated with the stronger network magnetic field. Moreover, 
recent results from \cite{ASEetal2017} find median values of $\sim60$~G from Ca II 8542~{\AA} observations of 
plage regions, although their estimated distribution of values has a long tail to larger values. 

The larger propagation speed measurements here typically have larger errors, which may 
account for some of the apparent spread in the distribution of speeds. An alternative may be that there are 
standing modes supported 
by the IN fibrils, which would lead to spurious measurements of fast propagation speeds. Similar results were obtained
in spicule measurements (\citealp{OKADEP2011}) and are still unexplained.

\medskip 
Finally, combining the measured wave amplitudes with propagation speeds, we are able to apply 
current solar magneto-seismology theory to the observations. We do so bearing in mind that
the theory is quite conservative in terms of what is assumed about the behaviour of the local 
plasma (Section~\ref{sec:seis}). The data reveals that the amplitude of the waves 
is non-constant along the waveguides, showing evidence of amplification and damping as a 
function of distance. The amplification is readily explained in terms of longitudinal stratification
of the local plasma, combining together changes in internal and ambient plasmas. Observations and simulations
suggest fibrils rise out of the surrounding 'bulk chromosphere' and protrude some way into the corona (leading
in part to a larger column mass), hence, we expect that there
will be large changes in the external density as the ambient atmosphere rapidly transitions from
chromosphere to corona. This situation would lead to the ambient density variations dominating 
the observed variation in amplitude, and masking the influence of internal density stratification. 
However, at present, it is not clear how to disentangle the two density profiles from the observed 
variation. In addition,
it is expected that the kink waves are subject to damping of some form, which will also
be entangled in the observed amplitude profile. Some of the profiles shown in Figure~\ref{fig:seisprof}
support the idea of the changes in external plasma dominating their behaviour, with measured variations 
in amplitude suggesting density scale-heights similar to that of 
a vertical gravitationally stratified atmosphere.\\

In conclusion, we provide here the first study of internetwork fibrils. The IN fibrils display
a pattern of repeated H$\alpha$ visibility suggesting repetitive heating events along a quasi-stable
magnetic flux tube. We also find evidence of flows along the longitudinal axis of the
fibril and observe quasi-periodic transverse propagating displacements that we interpret as the MHD kink 
wave. We suggest that both the flows and waves indicate the correspondence of an underlying magnetic field with 
the absorption feature visible in H$\alpha$.

\begin{acknowledgements}
KM acknowledges support from Northumbria University Research Development Fund and the Royal Astronomical Society.
RM is grateful to the Leverhulme Trust for the award of an Early Career 
Fellowship. All authors acknowledge IDL support provided by the Science \& Technologies Facilities Council. 
The Swedish 1-m Solar Telescope is operated on the island of La Palma by the Institute for Solar Physics (ISP) of Stockholm University
in the Spanish Observatorio del Roque de los Muchachos of the Instituto de Astrof\'isica de Canarias.
The observations were taken as part of work supported by the SOLARNET project (www.solarnet-east.eu), 
funded by the European Commissions FP7 Capacities Program under the Grant Agreement 312495.
\end{acknowledgements}

\bibliographystyle{aa}

\begin{table*}
\centering
\caption{\label{tab:one}Measured fibril properties.}
\begin{tabular}{lccccccc}
\hline\hline
{Index} & {$\xi$ (km)}  & {$P$ (s)} & {$v$ (km/s)} 
& {$C_{k}$ (km/s)}  & {Scale height} & {av no.} & Plot legend\\
\hline

1  & 33$\pm$1  & 130$\pm$1  & 2.59$\pm$0.08 & 961$\pm$307 & -842$\pm$1014& 10 & blue\\ 
2  & 28$\pm$1  & 107$\pm$2  & 2.57$\pm$0.16 & 405$\pm$252 & 593$\pm$933 & 8 & green\\ 
3  & 23$\pm$2  & 118$\pm$3  & 1.30$\pm$0.11 & 51$\pm$11 & - & 5 & -\\
4  & 86$\pm$2  & 109$\pm$1  & 8.70$\pm$0.23 & 89$\pm$12 & -  & 6 & - \\ 
5  & 53$\pm$1  & 67 $\pm$1  & 7.67$\pm$0.21 & 960$\pm$353 &-2232$\pm$9903& 8 & brown\\ 
6  & 54$\pm$2  & 80 $\pm$1  & 7.01$\pm$0.29 & 622$\pm$275 & -276$\pm$251&5 & pink \\ 
7  & 86$\pm$1  & 134$\pm$1  & 6.82$\pm$0.12 & 788$\pm$236 &  - &9 & - \\ 
8  & 82$\pm$1  & 103$\pm$1  & 8.83$\pm$0.12 & 303$\pm$61 & -4613$\pm$23738&5 & orange \\ 
9  & 118$\pm$1 & 123$\pm$1  & 11.08$\pm$0.12 & 939$\pm$111 & - &9 & -\\ 
10 & 118$\pm$1& 123$\pm$1  & 11.08$\pm$0.12 & 400$\pm$41 & 475$\pm$173&9 & gold\\
11 & 102$\pm$2 & 273$\pm$4  & 3.94$\pm$0.09 & 161$\pm$19 & 692$\pm$722&9 & red \\ 
12 & 171$\pm$3 & 180$\pm$1  & 10.90$\pm$0.22 & -489$\pm$218 & 172$\pm$62 &5 & orange\\ 
13 & 169$\pm$3 & 144$\pm$2  & 13.06$\pm$0.31 & 100$\pm$17&  - &5 & -\\ 
14 & 82$\pm$3  & 90$ \pm$1  & 5.24$\pm$0.21 & -732$\pm$197 & -2103$\pm$10772&6 & blue \\ 
15 & 74$\pm$1  & 114$\pm$1  & 7.34$\pm$0.09 & -677$\pm$35 & 601$\pm$176& 11 & gold\\ 
16 & 186$\pm$3 & 162$\pm$2  & 13.30$\pm$0.26 & 749$\pm$141 &  - &7 & - \\ 
17 & 60$\pm$1  & 106$\pm$1  & 6.36$\pm$0.13 & -845$\pm$382 & -380$\pm$271&5 & green \\ 
18 & 51$\pm$2  & 81 $\pm$1  & 7.15$\pm$0.21 & 325$\pm$41 &  -&10 & -\\ 
19 & 62$\pm$1  & 88 $\pm$1  & 8.40$\pm$0.22 & 249$\pm$29 &  - &9 & -\\ 
20 & 69$\pm$2  & 192$\pm$5  & 3.81$\pm$0.16 & 78$\pm$15 &  - &5 & -\\ 
21 & 79$\pm$1  & 108$\pm$1  & 6.62$\pm$0.09 & 1077$\pm$178 & 552$\pm$98&14 & purple \\ 
22 & 100$\pm$1 & 153$\pm$1  & 7.12$\pm$0.07 & -701$\pm$46 & -3688$\pm$2906 &15 & brown\\
23 & 78$\pm$2  & 157$\pm$2  & 5.52$\pm$0.17 & 55$\pm$8 &  - &6 & -\\ 
24 & 153$\pm$2 & 180$\pm$1  & 9.38$\pm$0.11 & 127$\pm$21 & - &9 & -\\ 
25 & 31$\pm$2  & 95 $\pm$1  & 2.99$\pm$0.16 & -139$\pm$35 & 128$\pm$102&5 & red \\ 
26 & 94$\pm$1  & 144$\pm$1  & 7.42$\pm$0.13 & -214$\pm$38 & -235$\pm$134&6 & pink \\ 
27 & 71$\pm$2  & 105$\pm$1  & 7.65$\pm$0.18 & -171$\pm$20 & 70$\pm$9&7 & purple\\ 
28 & 75$\pm$1  & 122$\pm$1  & 6.49$\pm$0.14 & 85$\pm$5 & 69$\pm$7 &8 & cyan\\ 
\end{tabular}
\tablefoot{Uncertainties on displacement and periods have been rounded to nearest integer value, or rounded up to 1
if less than 0.5.}
\end{table*}

\end{document}